\documentclass{article}

\usepackage[inline]{enumitem}
\usepackage[export]{adjustbox}
\usepackage{natbib, amsfonts,amsmath,graphicx,array,pifont,soul,amssymb,booktabs,multicol,multirow,setspace,euscript,verbatim,xspace,xcolor,standalone,pgfplots,pgf,comment,url,floatrow,algorithm,algpseudocode,float,listings,adjustbox,wrapfig,float,xifthen,etoolbox,microtype}
\usepackage[caption=false]{subfig}
\usepackage[colorlinks=true, allcolors=blue,backref=page]{hyperref}
\usepackage[short,c2]{optidef}
\usepackage[nameinlink]{cleveref}
\definecolor{lime}{HTML}{A6CE39}
\DeclareRobustCommand{\orcidicon}{%
	\begin{tikzpicture}
	\draw[lime, fill=lime] (0,0) 
	circle [radius=0.16] 
	node[white] {{\fontfamily{qag}\selectfont \tiny ID}};
	\draw[white, fill=white] (-0.0625,0.095) 
	circle [radius=0.007];
	\end{tikzpicture}
	\hspace{-2mm}
}

\foreach \x in {A, ..., Z}{%
	\expandafter\xdef\csname orcid\x\endcsname{\noexpand\href{https://orcid.org/\csname orcidauthor\x\endcsname}{\noexpand\orcidicon}}
}
\setlist[enumerate]{label=(\roman*.)}

\newcommand{\code}[1]{{\texttt{#1}}}
\renewcommand{\bar}{\overline}
\renewcommand{\tilde}{\widetilde}

\colorlet{rosso}{red!90!white}
\definecolor{green}{RGB}{45, 204, 53}
\definecolor{blue}{RGB}{59, 42, 247}

\allowdisplaybreaks
\crefname{lemma}{Lemma}{Lemmata}
\crefname{theorem}{Theorem}{Theorems}
\crefname{claim}{Claim}{Claims}
\crefname{proposition}{Proposition}{Propositions}
\crefname{algorithm}{Algorithm}{Algorithms}
\crefname{equation}{}{}
\crefname{definition}{Definition}{Definition}
\crefname{Cla}{Claim}{Claim}
\crefname{corollary}{Corollary}{Corollaries}
\crefname{remark}{Remark}{Remarks}
\crefname{example}{Example}{Examples}
\crefname{figure}{Figure}{Figures}
\crefname{section}{Section}{Sections}
\crefname{table}{Table}{Tables}
\crefname{enumi}{Statement}{Statements}
\crefname{line}{Line}{Lines}

\newcommand{\RBG}{\emph{RBG}\xspace}
\newcommand{\RBGs}{\emph{RBG}s\xspace}

\newcommand{\NASPs}{\emph{NASP}s\xspace}

\newcommand{\LCPs}{\emph{LCP}s\xspace}
\newcommand{\IPG}{\emph{IPG}\xspace}
\newcommand{\IPGs}{\emph{IPG}s\xspace}

\newcommand{\MPG}{\emph{MPG}\xspace}
\newcommand{\MPGs}{\emph{MPG}s\xspace}

\newcommand{\xminusi}{x^{-i}}
\newcommand{\xofi}{x^i}

\newcommand{\tildesigmaofi}{\tilde{\sigma}^i}

\newcommand{\XSetofi}{\mathcal{X}^i}

\newcommand{\clconvi}{\cl \conv (\XSetofi)}

\newcommand{\conv}{\operatorname{conv}}

\newcommand{\cl}{\operatorname{cl}}

\definecolor{commentstyle}{HTML}{45d14e}
\definecolor{codegray}{rgb}{0.5,0.5,0.5}
\definecolor{stringstyle}{HTML}{5ea2eb}
\definecolor{backcolour}{HTML}{f2f7ff}
\definecolor{keywordstyle}{HTML}{d63dfc}

\lstdefinestyle{ZERO}{
  backgroundcolor=\color{backcolour},
  commentstyle=\color{commentstyle},
  keywordstyle=\color{keywordstyle},
  numberstyle=\tiny\color{codegray},
  stringstyle=\color{stringstyle},
  basicstyle=\ttfamily\scriptsize,
  breakatwhitespace=false,
  breaklines=true,
  captionpos=b,
  keepspaces=true,
  numbers=left,
  numbersep=10pt,
  showspaces=false,
  showstringspaces=false,
  showtabs=false,
  tabsize=2
}
\newcommand{\cpp}{C\nolinebreak\hspace{-.05em}\raisebox{.4ex}{\tiny\bf +}\nolinebreak\hspace{-.10em}\raisebox{.4ex}{\tiny\bf +}\xspace}

\makeatletter
\newcommand{\xblue}[1][\@nil]{%
  \def\tmp{#1}%
   \ifx\tmp\@nnil
       {\color{blue} x^1}
    \else
       {\color{blue} x^1_#1}
    \fi}
\newcommand{\xred}[1][\@nil]{%
  \def\tmp{#1}%
   \ifx\tmp\@nnil
       {\color{red} x^2}
    \else
       {\color{red} x^2_#1}
    \fi}
\makeatother

 \bibpunct[, ]{(}{)}{,}{a}{}{,}%
 \renewcommand*{\backref}[1]{}
\renewcommand*{\backrefalt}[4]{%
    \ifcase #1 (Not cited.)%
    \or        (Cited on page~#2.) %
    \else      (Cited on pages~#2.) %
    \fi}
\pgfplotsset{compat=1.16}

\begin{document}

\newtheorem{theorem}{Theorem}
\newtheorem{proposition}{Proposition}
\newtheorem{definition}{Definition}
\newtheorem{remark}{Remark}
\newtheorem{example}{Example}
\newtheorem{lemma}[theorem]{Lemma}

\title{ZERO: Playing Mathematical Programming Games}

\author{
Gabriele Dragotto\footnote{\orcidA{} DS4DM, Polytechnique Montr\'eal \texttt{gabriele.dragotto@polymtl.ca}}, 
Sriram Sankaranarayanan\footnote{\orcidB{} P\&QM, Indian Institute of Management, Ahmedabad}, \\
Margarida Carvalho\footnote{\orcidC{} CIRRELT \& DIRO, Universit\'e de Montr\'eal},
Andrea Lodi\footnote{\orcidD{} DS4DM and Jacobs Technion-Cornell Institute, Cornell University }
}
\date{}
\maketitle     
\begin{abstract}
We present \emph{ZERO}, a modular and extensible \cpp library interfacing Mathematical Programming and Game Theory. ZERO provides a comprehensive toolkit of modeling interfaces and algorithms for Reciprocally Bilinear Games (\RBGs), i.e., simultaneous non-cooperative games where each player solves a mathematical program with a linear objective in the player's variable and bilinear in its opponents' variables. This class of games generalizes the classical problems of Operations Research to a multi-agent setting. 
ZERO modular structure gives users all the elementary ingredients to design new game-theoretic models and algorithms for \RBGs, and find their Nash equilibria. The library provides additional extended support for integer non-convexities, linear bilevel problems, and linear equilibrium problems with equilibrium constraints.
We provide an overview of the software's key components and showcase a Knapsack Game, i.e., a game where each player solves a binary knapsack problem. Aiming to boost practical methodological contributions at the interplay of Mathematical Programming and Game Theory, we release ZERO as open-source software. Source code, documentation and examples are available at \url{www.getzero.one}.
\end{abstract}      \bibliographystyle{plainnat}

\section{Why Games and Equilibria?}
The pioneering book from \citet{morgenstern_theory_1953} and the seminal papers from \citet{nash_equilibrium_1950,nash_noncoop_1951} transformed the scientific perspective on strategic behavior. The ubiquitous concepts of Nash equilibrium and rationality are now cornerstone concepts in Game Theory, with applications ranging from Economics to Social Sciences. The growing interest in game dynamics in the Operations Research community reflects a need to extend classical decision-making frameworks to multi-agent settings that can account for interactions among multiple decision-makers. The community devoted particular interest -- to name a few -- to bilevel programming (e.g., \citet{hu_using_2007,chinneck_branch-and-cut_2009,labbe2013bilevel,Caprara2016,fischetti_new_2017,Basu2018a,kleinert_survey_2021}) and its application in electricity markets and network pricing problems \citep{feijoo2015bilevel,Labbe1998,Brotcorne2001}, equilibrium problems with equilibrium constraints \citep{luo_mathematical_1996,WNMS1}, and more recently to integer programming games \citep{vaz_existence_2018,Carvalho2020computing,Dragotto_2021_CNP,cronert_equilibrium_2021,guo_copositive_2021,koppe_rational_2011,Dragotto_2021_ZERORegrets}. On the one hand, such empowering modeling capabilities unquestionably offer a tempting opportunity for extending the domain of influence of Operations Research. Arguably, multi-agent optimization frameworks can help provide enhanced models by contemplating the interactions decision-makers often take by pondering the influence of other stakeholders (e.g., other players). Additionally, they can help embed socially-beneficial outcomes by enlightening the nature of interaction among selfish decision-makers. For instance, \citet{WNMS1} provide insights on the role of a carbon tax in competitive international energy markets, \citet{carvalho_nash_2017} prove that the most rational outcome in their cross-border kidney exchange maximizes the social welfare (e.g., the sum of the objectives of all players). 
On the other hand, multi-agent models are as helpful as one can efficiently compute equilibria (or equivalent solution paradigms), thus highlighting the importance of theoretical and practical contributions for computing them. We believe that free and open-source software can foster experimentation in both practitioners' and researchers' communities, and hopefully lead to novel methodological advancements in the field.

\subsection{Background. }In this context, we introduce \emph{ZERO}, a modular \cpp package to handle Reciprocally-Bilinear Games (\RBGs), a special class of Mathematical Programming Games (\MPGs). An \MPG is a simultaneous game among $n$ players, each of which solves a mathematical program whose objective function is parametrized in other players' variables, and whose feasible region's description does not include other players' variables.
Although \MPGs are also Nash equilibrium problems ($NEPs$) \citep{facchinei_finite-dimensional_2003}, the \MPGs taxonomy we propose follows three assumptions: \begin{enumerate*} \item a set of constraints, for instance, a set of linear constraints and integer requirements, represent each player's moves. This set may be unbounded, contain infinitely or finitely many elements, and generally does not have a special structure. We do not assume the players' feasible sets to be continuous (i.e., in contrast to most of the $NEPs$ literature), nor that computing equilibria necessarily requires the solution of a complementarity problem, \item we aim to build a language intersecting both elements of Game Theory and Mathematical Programming, \item we aim to preserve the structure that constraints give to each player's problem. For instance, we may not drop any constraints to simplify the game without damaging its modeling capability.\end{enumerate*} For the above three reasons, we introduce the class of \MPGs to represent a wide variety of games among optimization problems.

ZERO provides support for a fundamental class of \MPGs, namely the class of \RBGs. Let the operator $\left( \cdot \right)^{-i}$ define $\left( \cdot \right)$ except $i$; e.g., if $x=(x^1,\dots,x^n)$, then $x^{-2}=(x^1,x^3,\dots,x^n)$.

\begin{definition}[Reciprocally-Bilinear Game \citep{Dragotto_2021_CNP}]
A \emph{Reciprocally-Bilinear Game} (\RBG) is an \MPG among $n$ players, where each player $i=1,2,\dots,n$ solves the optimization problem
\begin{mini!}
    {\xofi}{f^i(x^i,x^{-i}) = (c^i) ^\top \xofi + (\xminusi)^\top C^i\xofi  \protect\label{eq:RBG:Obj}}
    {\label{eq:RBG}}{}
    \addConstraint{\xofi}{\in \XSetofi \protect\label{eq:RBG:Constraints}}
\end{mini!}
where $\XSetofi \subseteq \mathbb{R}^{m_i}$, and $C$ and $c$ are a matrix and a vector of appropriate dimensions, respectively. An \RBG is \emph{polyhedrally-representable} if $\clconvi$ is a polyhedron for each $i$, and one can optimize a linear function over each $\XSetofi$. 
\label{def:RBG}
\end{definition}

In \RBGs, the $i$-th player objective function $f^i(x^i,x^{-i})$ -- or payoff function for $i$ -- is linear in $x^i$ and contains bilinear products with $x^i$ and $x^{-i}$. Further, since \RBGs are \MPGs, the description of each player's feasible region $\XSetofi$ does not contain other players' variables, and the $i$-th player optimization problem is \emph{parametrized} in $x^{-i}$, namely plugging $x^{-i}$ as a parameter results in an optimization problem purely in the variables $x^i$. When $n=1$, the \RBG in \cref{def:RBG} is a single optimization problem in $x^i$. Whenever $n>1$, \RBGs become expressive models extending typical Operations Research tasks -- such as resource allocation, scheduling, or routing -- to a multi-agent setting.  Consider, for instance, the emblematic $0/1$ Knapsack Problem; given a set of items, a decision-maker selects some of them to maximize the sum of the profits associated with each item, subject to a capacity constraint. A multi-agent extension of this problem is the so-called Knapsack Game as in \cref{def:KPG}, where $n$ players simultaneously solve a $0/1$ Knapsack Problem.

\begin{example}[Knapsack Game]
A Knapsack Game is an \RBG where each player solves the optimization problem
\begin{align}
    \max_{x^i} \{ (c^i) ^\top \xofi + (\xminusi)^\top C^i\xofi : (a^i)^\top x^i \le b^i, x^i \in \{0,1\}^{m_i} \}
\end{align}
where $m_i$ is the number of items for player $i$, $b^i \in \mathbb{Z}$, $a^i \in \mathbb{Z}^{m_i}$, $c^i\in \mathbb{Z}^{m_i}$, and $C^i$ is an integer-valued matrix of appropriate size.
\label{def:KPG}
\end{example}

In this game, player $i$ has not only to consider a feasible packing of items maximizing the profits associated with the vector $c^i$, but has to look out for the positive or negative impact of the interaction of its packings with the ones of its opponents (the $C^i$ products). Besides being an \RBG, the Knapsack Game is also an Integer Programming Game (\IPG), namely an \MPG where each player solves a mixed-integer problem \citep{koppe_rational_2011}.
The sets $\clconvi$ are the so-called integer hulls associated with each player's $0/1$ knapsack polytope, and each point $\bar{x}^i \in \XSetofi$ is a \emph{pure-strategy} for $i$, namely a solution to the knapsack problem for $i$. In general, each $\tildesigmaofi \in \clconvi$ is a \emph{mixed-strategy}, namely a point inside the $0/1$ knapsack polytope. The central question is then to determine what is a solution to the above game. In an optimization problem, we usually search for an optimal solution that maximizes (minimizes) the objective function while fulfilling the constraints. However, in a game, a solution should be \emph{stable}, meaning that it should be mutually optimal for all the players, and not only a subset of them. The most famous solution paradigm in Game Theory is the one of Nash Equilbirum, a solution where each player cannot \emph{unilaterally} deviate from it while improving its payoff. We formally define the concept of Nash equilibrium for \RBGs in \cref{def:NE}; we remark that in \cref{def:RBG} players are minimizing their objective functions, and improving a payoff means decreasing it.

\begin{definition}[Pure-Strategy Nash Equilibrium]
A strategy profile $\bar{x}=(\bar{x}^1,\dots,\bar{x}^n)$ is a \emph{Pure-Strategy Nash Equilibrium} for an \RBG as in \cref{def:RBG} if, for each player $i$ and strategy $\tilde{x}^i \in \XSetofi$, then $f^i(\bar{x}^i,\bar{x}^{-i}) \le f^i(\tilde{x}^i,\bar{x}^{-i})$. 
\label{def:NE}
\end{definition}

In other words, at the equilibrium $\bar{x}=(\bar{x}^1,\dots,\bar{x}^n)$, no player $i$ can possibly pick a strategy $\tilde{x}^i \neq \bar{x}^i$ so that $f^i(\bar{x}^i,\bar{x}^{-i})>f^i(\tilde{x}^i,\bar{x}^{-i})$. In this sense, the equilibrium strategy is resilient to the moves of each player's opponents and provides a mutually-optimal solution. The \emph{Mixed-Strategy Nash equilibrium} relaxes the definition of Pure-Strategy Nash equilibrium by allowing players to select not only pure-strategies, but in general mixed-strategies.

\section{Our Contributions} 
ZERO provides advanced and modular \cpp toolkits to formulate \RBGs and compute their Nash equilibria, with high-level APIs for practitioners and low-level ones for researchers and experienced users. We summarize the most important contributions as follows.
\begin{enumerate}

\item ZERO is the first library to support non-cooperative simultaneous games where players solve mathematical programs. Other Game Theory solvers, such as Gambit \citep{gambit} only support finite games in normal form (games with finitely many players, finitely many strategies and outcomes).

\item The library has a modular structure designed for allowing extensibility. Each component -- or \emph{module} -- independently performs a specific task and interacts with the others through well-defined interfaces. For instance, the natively embedded algorithms interface with the base modules allowing the development of sophisticated computational routines. Users can either use the included algorithms or implement custom ones depending on the desired level of control.

\item The library is an abstract layer bridging typical Mathematical Programming and Game Theory and focuses on the interaction and orchestration among external libraries and native modules. We delegate most of the standard mathematical programming routines to specialized software, thus integrating popular and well-maintained tools available in the Operations Research community. For instance, we solve mathematical programs through Gurobi \citep{gurobi} and PATH \citep{path}, we generate cutting planes with Coin-OR Cgl \citep{coinor}, and we perform linear algebra operations through Armadillo \citep{armadillo}. 

\item ZERO can work as an off-the-shelf solver for \RBGs without the need for a deep technical understanding of the algorithmic details. We provide a series of high-level interfaces designed specifically for some classes of \RBGs, along with standardized instance file schemes and plug-and-play shell executables. On the one side, ZERO provides high-level APIs for practitioners and industrial parties to experiment with our high-level APIs. On the other side, we target experienced users by offering advanced tools to build sophisticated models and algorithms.
\end{enumerate}

\section{Overview}
We briefly give an overview of ZERO: the detailed documentation for the software is available online at \url{www.getzero.one}. Our library currently supports any polyhedrally-representable \RBG, and further provides additional tools (i.e., high-level modeling APIs) for two specific types of games. First, \IPGs, namely \MPGs where each player solves an integer program; in particular, ZERO supports \IPGs that are also \RBGs, and hence have a bilinear objective as in \cref{def:RBG}. Second, Nash games Among Stackelberg Players (\NASPs), a class of Equilibrium Problems with Equilibrium Constraints among the leaders of continuous bilevel games \citep{WNMS1}.

\begin{figure}[!ht]
    \includegraphics[width=0.99\textwidth,valign=t]{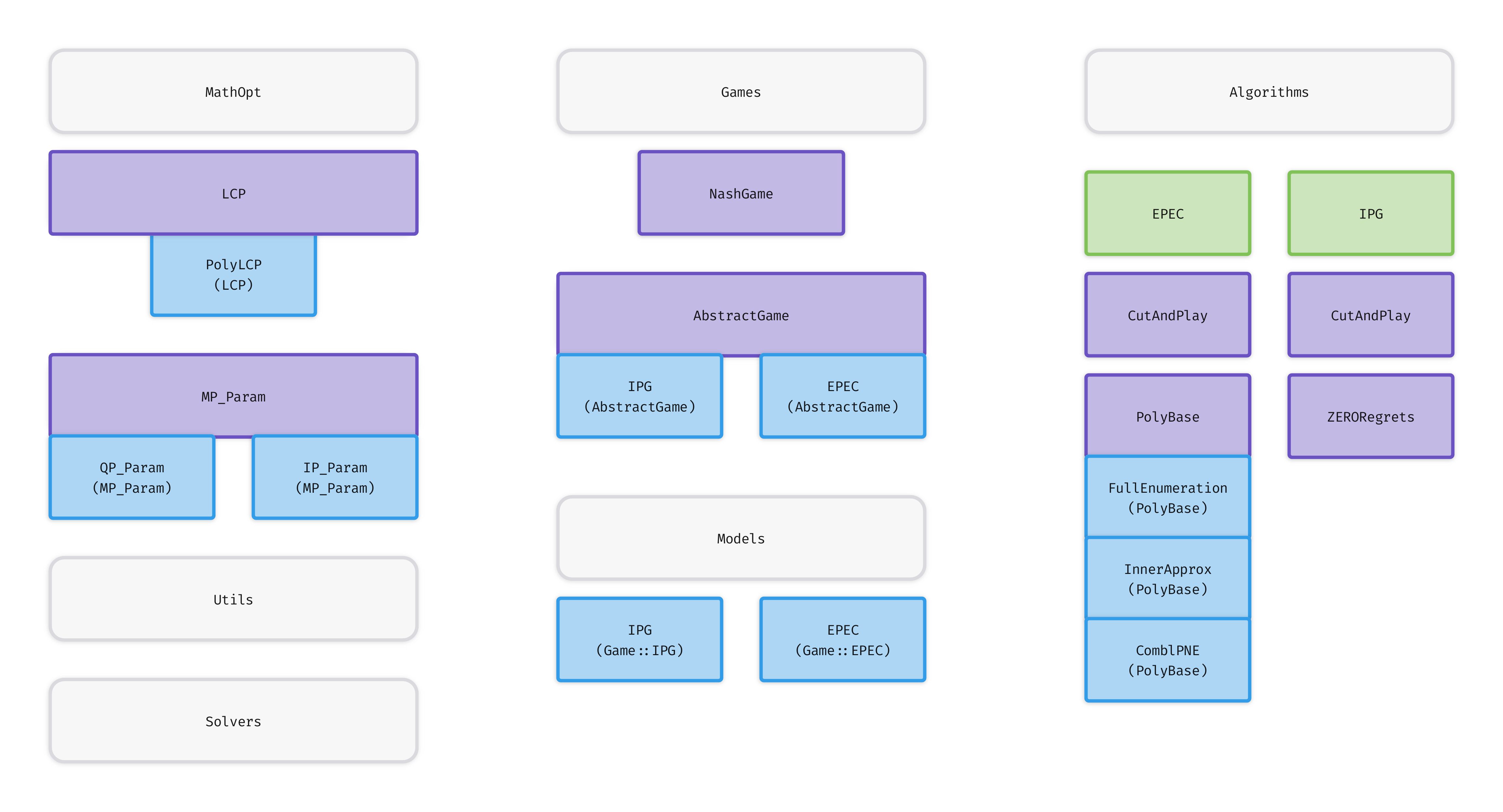}
    \caption{A schematic view of ZERO's modules, 10000 lines of code, 50 files, 40 classes, and 450 functions. The \emph{namespaces} are in gray, and the relative content is grouped below. The primitive classes are in purple, and the associated inheritor classes are in blue. Nested namespaces are in green.}
     \label{fig:Classes} 
\end{figure}

\paragraph{Modules and Namespaces. }ZERO's modules are classes defined inside a suitable \emph{namespace},  namely a larger scope grouping modules with similar functions or goals. In the sequel, we provide an overview of the software architecture.
The namespace \code{MathOpt} contains the necessary optimization tools for defining and solving mathematical programs -- for instance, \code{MathOpt::IP\_Param} for parametrized mixed-integer linear programs, and \code{MathOpt::LCP} for linear complementarity problems (\LCPs) -- as well as helper functions (e.g., \code{MathOpt::convexHull} for computing the convex hull of a union of polyhedra). This class provides a layer between ZERO and the external solvers such as Gurobi and PATH.
Arguably, the most relevant namespace is the one of \code{Games}, which implements the abstraction of specific \RBGs, such as \code{Games::IPG} for \IPGs, and \code{Games::EPEC} for \NASPs. The modules inside this namespace orchestrate a tight integration among all the other modules and provide several low-level APIs to the user. The namespace \code{Algorithms} contains the algorithms to compute the Nash equilibria for \RBGs. Such algorithms are inside the modules of this namespace and closely coordinate with the modules in \code{Games}; for instance, the class \code{Algorithms::IPG::CutAndPlay} associated with the Cut-And-Play algorithm for \IPGs and \NASPs \citep{Dragotto_2021_CNP} coordinates with both \code{Games::EPEC} and \code{Games::IPG}.
Other than advanced users, ZERO aims to target practitioners that may only be interested in plug-and-play usage of the software. Thus, in the namespace \code{Models} we provide high-level APIs allowing users to quickly model and solve off-the-shelves instances of \IPGs and \NASPs. Furthermore, we propose a standardized format for instances encoded through the data-interchange format \code{JSON} \citep{json}, and integrate complementary helper functions to manage the input and output files.
We also include two shell executables working with standardized instance formats allowing users to deploy the algorithms and solve instances on the run.
Finally, the namespace \code{Utils} provides some simple helper functions for writing and reading files, as well as additional numerical and linear algebra utilities. \cref{fig:Classes} provides a schematic representation of the architecture.

\section{Modeling the Knapsack Game}
We showcase how to model an instance of the Knapsack Game of \cref{def:KPG} with ZERO. Let \emph{blue} be Player 1 and \emph{red} be Player 2. Each player $i$ seeks to pack $m_i=2$ items into its knapsack with capacity $b^i=5$. The optimization problems for blue and red are in \cref{eq:KP1} and \cref{eq:KP2}, respectively.

\begin{minipage}[t]{.45\textwidth}
\vspace{-0.2cm}
\raggedright
\vspace{-0.5cm}
\begin{maxie}
    {\xblue}{\xblue[1] + 2\xblue[2] - 2\xblue[1]\xred[1] -3\xblue[2]\xred[2]  \protect\label{eq:KP1:Obj}}
    {\label{eq:KP1}}{}
    \addConstraint{3\xblue[1]+4\xblue[2]}{\le 5, \xblue \in \{0,1\}^2 \protect\label{eq:KP1:Constraint1}}
\end{maxie}
\end{minipage}
\hfill \vline \hfill
\begin{minipage}[t]{.45\textwidth}
\vspace{-0.2cm}
\raggedleft
\vspace{-0.5cm}
\begin{maxie}
    {\xred}{3\xred[1] + 5\xred[2] - 5\xred[1]\xblue[1] -4\xred[2]\xblue[2]  \protect\label{eq:KP2:Obj}}
    {\label{eq:KP2}}{}
    \addConstraint{2\xred[1]+5\xred[2]}{\le 5, \xred \in \{0,1\}^2 \protect\label{eq:KP2:Constraint1}}
\end{maxie}
\end{minipage}
\\

\noindent This problem has 3 Nash equilibria: the Pure-Strategy Nash equilibria
$(\xblue[1],\xblue[2],\xred[1],\xred[1]) = (0, 1, 1, 0)$, $(\xblue[1],\xblue[2],\xred[1],\xred[1]) = (1, 0, 0, 1)$, and the Mixed-Strategy Nash equilibrium $(\xblue[1],\xblue[2],\xred[1],\xred[1]) = (\frac{2}{9}, \frac{7}{9}, \frac{2}{5}, \frac{3}{5})$.
We attempt to find one of them by using the \emph{Cut-And-Play} algorithm from \citep{Dragotto_2021_CNP}. Intuitively, this algorithm iteratively refines each players' feasible region starting from its linear relaxations (i.e., the polyhedron given by dropping the integrality constraint in either \cref{eq:KP1} or \cref{eq:KP2}). Specifically, the algorithm iteratively refines the linear relaxations adding cutting planes (some of which generated by Cgl from Coin OR \citep{coinor}) or by branching until it finds a Nash equilibrium. 

\begin{figure}[!ht]
\centering
\begin{lstlisting}
#include <zero.h>                                       |\label{code:header}| 

int main(int argc, char **argv) {
	 GRBEnv GurobiEnv;                                     |\label{code:gurobi}| 
	 Models::IPG::IPGInstance IPG_Instance; // The IPG Instance     |\label{code:instance}| 
	 int                      numItems = 2, numPlayers = 2;
	 arma::vec    c(numItems); // Profits c in the objective            |\label{code:dataO1}| 
	 arma::sp_mat C( (numPlayers-1) * numItems, numItems); // C terms in the objective
	 arma::sp_mat a(1, numItems); // LHS for Knapsack constraint
	 arma::vec    b(1); // RHS for constraints
	 arma::vec IntegerIndexes(numItems); // The index of the integer variables
	 VariableBounds VarBounds = {{0, 1}, {0, 1}};   // Implicit bounds (LB,UB) on variables.  |\label{code:dataO2}| 

	 //Fill the values in the parameterized integer problem      |\label{code:datafill1}|   
	 b(0)    = 5; // Knapsack Capacity
	 for (unsigned int i = 0; i < numItems; ++i)
		IntegerIndexes.at(i) = i;

	 C(0, 0) = 2;  C(1, 1) = 3; // C terms in the objective for player Blue
	 a(0, 0) = 3;  a(0, 1) = 4; // Knapsack Constraints
	 c(0) = -1; c(1) = -2;   // The standard is minimization, hence minus           |\label{code:datafill2}| 
	   
	 // Create a parametrized Integer Program for player Blue
	 MathOpt::IP_Param PlayerBlue(C, a, b, c, IntegerIndexes, VarBounds, &GurobiEnv);    |\label{code:parametrized1}| 

	 // Parametrized Integer Program for player Red. |\label{code:repeat1}| 
	 C(0, 0) = 5; C(1, 1) = 4; a(0, 0) = 2; a(0, 1) = 5; c(0) = -3; c(1) = -5;                             

	 MathOpt::IP_Param PlayerRed(C, a, b, c, IntegerIndexes, VarBounds, &GurobiEnv);            |\label{code:repeat2}| 

	 // Add the players to the instance. We can also specify a file path to write the instance
	 IPG_Instance.addIPParam(PlayerBlue, "PlayerBlue_KP");                      |\label{code:add1}|       
	 IPG_Instance.addIPParam(PlayerRed,  "PlayerRed_KP");                       |\label{code:add2}| 
	 IPG_Instance.save("A_Knapsack_Game"); // Save the instance with the standardize format       |\label{code:save}|       
	 Models::IPG::IPG KnapsackGame(&GurobiEnv, IPG_Instance); // Create a model from the instance       |\label{code:solveStart}|       
	 // Select the equilibrium to compute a Nash Equilibrium
	 KnapsackGame.setAlgorithm(Data::IPG::Algorithms::CutAndPlay);                              |\label{code:setAlgo}|       
	 // A few optional settings                                                         |\label{code:opt1}|       
	 KnapsackGame.setDeviationTolerance(3e-4); // Numerical tolerance
	 KnapsackGame.setNumThreads(4); // How many threads, if supported by the solver?
	 KnapsackGame.setLCPAlgorithm(Data::LCP::Algorithms::MIP); // How do we solve the LCPs?
	 KnapsackGame.setTimeLimit(5); // Time limit in second
	 KnapsackGame.finalize(); // Lock the model                                             |\label{code:opt2}|  
	 // Run and get the results                                                 
	 KnapsackGame.findNashEq();                                                     |\label{code:solve}|  
	 KnapsackGame.getX().at(0).print("Player Blue:");  // Print the solution           |\label{code:sol1}|  
	 KnapsackGame.getX().at(1).print("Player Red:");                               |\label{code:sol2}|  

}

\end{lstlisting}
\caption{An Example of a \cpp instantiation of a 2-player Knapsack Game in ZERO}
\label{fig:KnapsackGame}
\end{figure}

\paragraph{Modeling and solving with ZERO. } \cref{fig:KnapsackGame} demonstrate the use of our high-level API for \IPGs by modeling the Knapsack Game in \cref{eq:KP1,eq:KP2}. We start by including the only header file \texttt{zero.h} in \cref{code:header} -- which contains the specifications for the entire library -- and by creating a new Gurobi environment in \cref{code:gurobi}. In \cref{code:instance} we create a new empty \IPG instance (\texttt{Models::IPG::IPGInstance}), which we will later populate with the programs in \cref{eq:KP1,eq:KP2}. From \cref{code:dataO1} to \cref{code:dataO2}, we create the objects holding the data for the integer programs, for instance, the vector $a$ for the knapsack constraint and the vector \texttt{IntegerIndexes} containing the indices of the integer-constrained variables. We fill in the data from \cref{eq:KP1} from \cref{code:datafill1} to \cref{code:datafill2}, and create the (parametrized) integer program for player \emph{blue} in \cref{code:parametrized1} with a constructor of \texttt{MathOpt::IP\_Param}. The latter class infers the number of parameters -- namely the number other players variables -- by counting the number of rows of $C^1$; in this case, the parameters are $2$, and they are associated to the choices of Player 2. From \cref{code:repeat1} to \cref{code:repeat2}, we iterate this data-filling process for \emph{red}, and eventually add the two parametrized integer programs to the \texttt{IPG\_Instance} in \cref{code:add1,code:add2}. In \cref{code:save}, we save the instance with the standardized data format for ZERO instances.
The solution process starts from line \cref{code:solveStart}, where we instantiate -- in the object \texttt{KnapsackGame} -- an \IPG model  with the data contained in \texttt{IPG\_Instance}. We employ the constructor of \texttt{Models::IPG::IPG} by also specifying a pointer to the Gurobi environment. In \cref{code:setAlgo}, we instruct ZERO to use the \emph{Cut-And-Play} algorithm to solve \texttt{KnapsackGame}. In \cref{code:opt1,code:opt2}, we set some extra options, and finally start computing the Nash equilibria in \cref{code:solve} by calling the method \texttt{Models::IPG::IPG::findNashEq()}. We print the Nash equilibrium found by the \emph{Cut-And-Play} in \cref{code:sol1,code:sol2}.

\section{Conclusions and Future Directions}
We introduced ZERO, a multi-purpose $C$++ library offering the base ingredients to help users model and solve \RBGs. On the one side, ZERO implements high-level and intuitive APIs to formulate \RBGs and solve them. On the other side, its modular and extensive design enables advanced users and researchers to build customized algorithms. 
A current limitation of ZERO is the availability of only two mathematical programming solvers. We plan to extend further the support for other solvers, such as SCIP \citep{scip}. Furthermore, we believe future methodological advancements will likely enable us to extend our support to other classes of \MPGs and \RBGs. Naturally, this is conditional to the development of the appropriate mathematical tools to do so. Indeed, we release ZERO with the ambition to foster methodological and applied research in this newly developing field at the intersection of Game Theory and Mathematical Programming.

    \section*{Acknowledgements}
     This work has been supported by the Canada Excellence Research Chair  in Real-time Decision-making (DS4DM).

\bibliography{Biblio.bib}
\label{sec:bib}

\end{document}